\documentclass[runningheads]{llncs}
\usepackage[T1]{fontenc}

\usepackage{newtxtext}        
\usepackage[varvw]{newtxmath}       
\usepackage{booktabs}
\usepackage{pifont}
\usepackage{amsmath}
\usepackage[table]{xcolor}
\usepackage{colortbl}
\usepackage{arydshln}
\usepackage[nolist]{acronym}
\usepackage{todonotes}
\usepackage{multirow}
\usepackage{graphicx}

\definecolor{mgreen}{HTML}{B4E5A2}
\definecolor{mred}{HTML}{F6C6AD}

\newcommand{\cmark}{\ding{51}}
\newcommand{\xmark}{\ding{55}}

\begin{document}

\begin{acronym}
  \acro{AI}{Artificial Intelligence}
  \acro{CLI}{Command Line Interface}
  \acro{CNN}{Convolutional Neural Networks}
  \acro{DoS}{Denial of Service}
  \acro{DL}{Deep Learning}
  \acro{DNN}{Deep Neural Network}
  \acro{FPR}{False Positive Rate}
  \acro{GUI}{Graphical User Interface}
  \acro{HERA 2.0}{Holistic nEtwork featuRes Aggregator}
  \acro{IDS}{Intrusion Detection System}
  \acro{IF}{Isolation Forest}
  \acro{ML}{Machine Learning}
  \acro{NID}{Network Intrusion Detection}
  \acro{NIDS}{Network Intrusion Detection Systems}
  \acro{RNN}{Recurrent Neural Networks}
  \acro{SDAID}{Signature and Deep Analysis-based Intrusion Detection}
  \acro{SOC}{Security Operations Center}
  \acro{SIEM}{Security Information and Event Management}
  \acro{XGB}{eXtreme Gradient Boosting}
\end{acronym}

\title{A Unified Framework for Intelligent Cyber Attack Detection and Alert Correlation}

\author{Miguel Silva\inst{1}\orcidID{0009-0008-6630-9939} \and
João Vitorino\inst{1}\orcidID{0000-0002-4968-3653} \and
Daniela Pinto\inst{1}\orcidID{0009-0000-3003-6694} \and
Eva Maia\inst{1}\orcidID{0000-0002-8075-531X} \and
Ivone Amorim\inst{1}\orcidID{0000-0001-6102-6165} \and
Isabel Praça\inst{1}\orcidID{0000-0002-2519-9859}}

\authorrunning{M. Silva et al.}

\institute{GECAD, School of Engineering of the Technical University of Porto, rua Dr. António Bernardino de Almeida, 4249-015 Porto, Portugal
\email{mdgsa@isep.ipp.pt}}

\maketitle              

\begin{abstract}
A Security Operations Center (SOC) plays a critical role in detecting, correlating, and responding to modern cyber threats. However, SOCs continue to face challenges related to alert overload, fragmented detection capabilities, and limited collaboration between analysts and automated systems. These issues often result in delayed incident response, analyst fatigue, and an increased risk of overlooking sophisticated, multi-stage attacks that unfold gradually across different layers of the network. To address these challenges, this paper proposes a unified framework for network traffic classification and alert correlation through the use of signature rules, machine learning, and sequential pattern mining algorithms. The solution was evaluated in a controlled testbed environment emulating realistic, multi-stage attack scenarios, including malware exploitation, lateral movement, reconnaissance, and data exfiltration. The obtained results demonstrate the effectiveness of the proposed framework in detecting and correlating complex attack behaviors, while providing analysts with actionable insights in line with a human-in-the-loop approach to SOC operations.

\keywords{Artificial intelligence  \and Data analytics \and Intrusion detection \and Network traffic analysis \and Cybersecurity.}
\end{abstract}

\section{Introduction}
The continuous growth of cyber threats to digital infrastructures has created a need for more sophisticated and intelligent systems to protect them. \ac{NIDS} and \ac{SIEM} have been continuously used to monitor network traffic, endpoint activities, and aggregate security events across networks~\cite{2025_shaukat_nids}. While \ac{NIDS} focus on identifying malicious activities in a computer network, \ac{SIEM} platforms centralize log management and correlation across multiple security tools and endpoints.

Although effective at their respective functions, they operate in isolation, failing to detect sophisticated attacks that exhibit complex patterns spanning multiple network layers~\cite{2022_kotenko_nids}. Signature-based \ac{NIDS} only identify known attack patterns being ineffective against zero-day exploits, while anomaly-based \ac{NIDS} detect novel threats, commonly using \ac{AI} algorithms at the cost of higher false alarm rates~\cite{2022_lata_nids}. These separated approaches result in missed detection of complex attacks and high volumes of false positives that affect \ac{SOC}'s effectiveness.

This work proposes a unified framework that performs an intelligent detection of cyber attacks and context-aware correlation of alert logs, aiming to provide more relevant insights for \ac{SOC} analysts. The proposed solution integrates three modules for network traffic collection, detection and classification of malicious traffic, and correlation of sequences of alerts across different sources through the use of computationally efficient \ac{ML} algorithms. The three modules were deployed and validated in a collaborative \ac{SOC} with real-world cyber attacks, providing detection capabilities through the combination of signature rule matching, anomaly detection, and multi-class classification.

This paper is organized as follows. Section~\ref{sec:related} summarizes the recent contributions from the research community regarding intelligent collaborative \ac{SOC}. Section~\ref{sec:solution} details the architecture and modules of the proposed solution. Section~\ref{sec:case} presents and discusses the case study in a collaborative \ac{SOC}. Finally, Section~\ref{sec:conclusion} presents the main conclusions and future work.

\section{Related Work}
\label{sec:related}

The deployment of \ac{NIDS} has progressed from single-method approaches to hybrid architectures that address the known limitations of traditional detection mechanisms. While early signature-based detection relied on predefined attack patterns and failed to detect unknown threats and zero-day attacks, anomaly-based detection could identify novel attacks~\cite{10763627}, but suffered from high false positive rates~\cite{2022_quiazzane_nids,10500368}. This is a fundamental trade-off, as signature-based detection is very precise but narrow, while anomaly-based detection can adapt to an environment and detect deviations, but is noisier~\cite{Hozouri2025}. These limitations can be tackled by the development of hybrid \ac{NIDS} architectures that leverage both signature-based and anomaly-based capabilities.

Over the years, different solutions have been proposed for integration in \ac{NIDS}. Chiba et al.~\cite{2019_chiba_nids} established a framework which combined Suricata as a signature-based detector using an \ac{IF} for anomaly detection. The authors were able to reduce detection times and computational overhead while maintaining accuracy when using signature capabilities before anomaly. Similarly, Quiazzane et al.~\cite{2022_quiazzane_nids} also integrated Suricata with \ac{ML}-based anomaly detection, however, this time with decision tree-based algorithms.

Other more complex integrations of \ac{ML} included the framework by Vo et al.~\cite{2022_vo_nids} with \ac{DL} techniques. These authors introduced the \ac{SDAID} algorithm which combines signature detection with an ensemble learning model of \ac{DNN} and \ac{XGB}. Vikram et al.~\cite{2024_vikram_nids} further demonstrated the effectiveness of \ac{DL} models by including \ac{CNN} and \ac{RNN}, which outperformed other \ac{ML} techniques.

To reduce the high volume of alerts and summarize them to provide more insights into multi-stage attack scenarios, several data mining algorithms have been adapted to \ac{NIDS}~\cite{Albasheer2022}. Since the alert logs are structured as a large time series, sequential pattern mining algorithms such as PrefixSpan~\cite{2001_prefixspan} are a promising approach for analyzing temporally ordered alerts. PrefixSpan can search alert sequences for recurring patterns and correlations, although it requires the normalization of alerts from heterogeneous sources over different time windows~\cite{Kotenko2022}.

Overall, previous studies have demonstrated that hybrid approaches can outperform single-method systems. However, to the best of our knowledge, there are no publicly available systems that integrate multiple signature, anomaly and classification approaches with sequence-based alert correlation.

\section{Proposed Framework}
\label{sec:solution}

The unified framework for cyber-attack detection and alert correlation is composed of three software modules that can be fully integrated in a collaborative \ac{SOC}: Collection, Detection, and Analytics. The first monitors one or more network interfaces to collect and decode network packets, aggregate them into network flows, and extract statistical features in a tabular data format. The second combines signature rule matching at a packet level with anomaly detection and multi-class classification at a flow level, performing the fusion of multiple types of alerts to record a single alert log with a combined severity level for each flow. Finally, the third extracts sequences of alert logs over multiple time windows and generates a context-aware correlation according to the type and source of each alert.

\begin{figure}[h!]
    \centering
    \includegraphics[width=\linewidth]{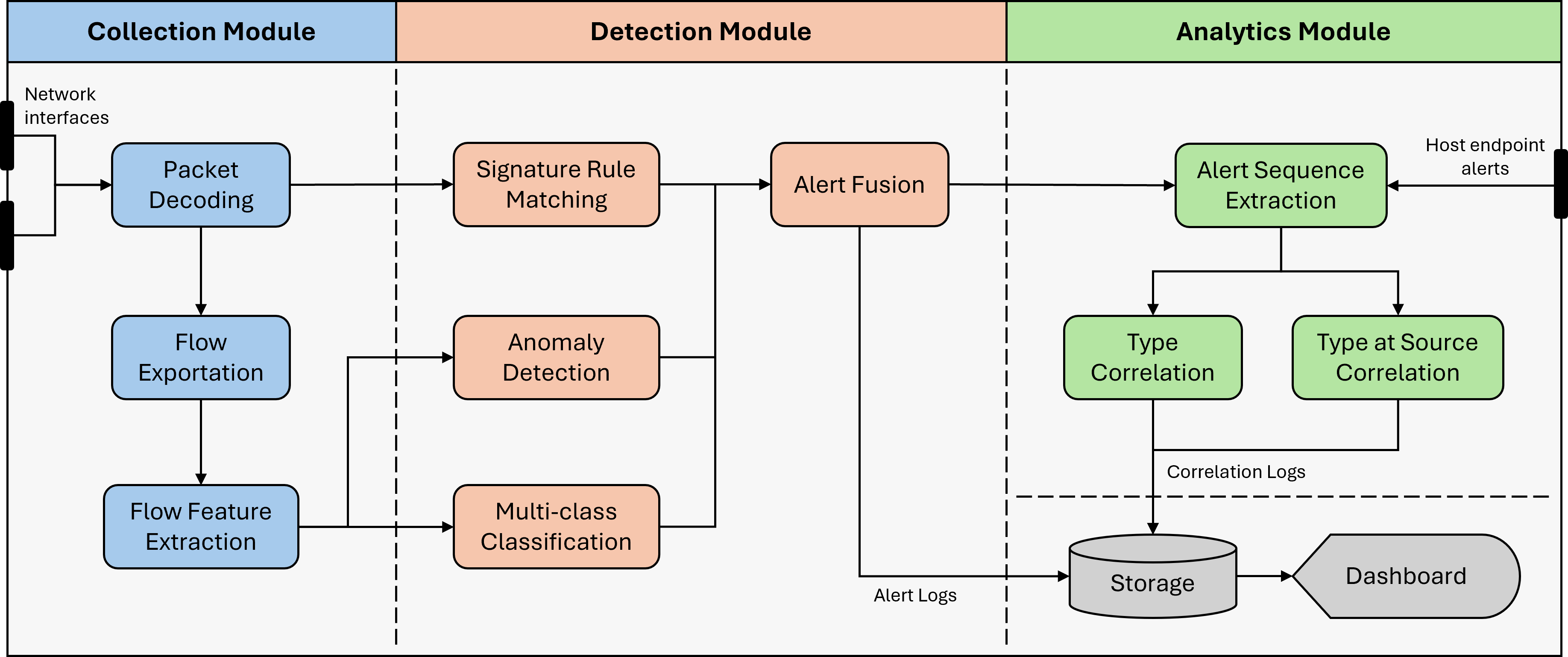}
    \caption{Hybrid detection and correlation architecture.}
    \label{fig:architecture}
\end{figure}

Figure~\ref{fig:architecture} provides an overview of the architecture, highlighting the flow of information from raw traffic at the network interfaces to processed alerts and correlations that are stored and provided in visualization dashboards. The raw traffic is obtained from one or more network interfaces being monitored, and the correlations can also include additional alert logs coming from each host's endpoint detection software. The following subsections provide detailed descriptions of each module.

\subsection{Collection Module}

The proposed solution has a dual stream of data to use for detection. Specifically, it processes two complementary types of data simultaneously, raw network packets and the exported flows from the same packets. The system utilizes the packets directly, while concurrently extracting flows and features from those same packets to enrich the detection process. These two data types provide distinct advantages and challenges. Packets allow for greater granularity but result in a higher computational cost due to their volume. Meanwhile, flows summarize packet information, drastically simplifying processing requirements while accepting a certain degree of information loss~\cite{2020_vormayr_tutorial}. By using both forms, the solution takes advantage of the best of the two data types.

\vspace{-1em}
\subsubsection{Network Packets.}
To capture the packets sent to the component, an \ac{IDS}, namely Suricata~\footnote{https://suricata.io/}, was deployed. It was also configured to save all communications, regardless of their protocol, to ensure the collection of all network traffic present on the network. Furthermore, Suricata, working as a packet collector, results in formatted outputs with statistical information of the packets captured, ready to be consumed by external components. For the proposed solution, the \ac{IDS} was set to export every packet's information in intervals of one minute.

\vspace{-1em}
\subsubsection{Network Flows.}
For the creation of flows, a flow exporter tool is necessary. These tools allow for the aggregation of packets into flows based on a set of values defined by the tool that performs the action. Commonly associated, the tool that aggregates the flows is also capable of extracting features based on the statistical information of the packets it uses to transform into flows. For the proposed solution, the flow exporter tool considered was \ac{HERA 2.0}~\cite{2024_pinto_HERA}~\footnote{https://github.com/danielaapp/HERA-2.0}, a tool that has been used to produce \ac{NID} datasets~\cite{2025_pinto_exporter,2025_silva_genis,2026_vitorino_flows}. This tool is designed to streamline the creation of flow-based datasets, allowing for its use with either \ac{CLI} or \ac{GUI}. Other than that, it also provides the possibility of using it in an offline or online mode. The tool listens on the network interfaces and generates flows that correspond to the packets of each connection over 60 seconds, extracting statistical features in a tabular data format for the following modules.

\subsection{Detection Module}

To identify cyber attacks in the network, the \ac{IDS} capabilities are used at a packet level to identify suspicious behavior. The \ac{IDS} utilizes the Open Emerging Threat Rules and custom rules to match packet statistics to detect known attacks, as well as identify possible threats, such as scans and bruteforce attempts, from either outside the network (internet) or from the inside, using a trusted machine. The results from the rule detection are added to the packet logs, specifically in the category, where a general category from the rule describes the detection, and the event type of the packet, which is set to alert if a rule was triggered. Furthermore, the rules are divided into three levels, 1 to 3, where the lowest level represents the most critical and severe attack detected.

On the flow level, two \ac{ML} models were trained to detect and identify suspicious traffic. The training dataset was built from traffic flows extracted from several datasets - GeNIS~\cite{2025_silva_genis}, HIKARI-2021~\cite{hikari21}, CICIDS-2017~\cite{cicids}, and UNSW-NB15~\cite{unsw} - using the HERA 2.0 flow exporter. These datasets were selected because they represent similar attacks generated through different architectures and captured at different points in time, allowing the models to generalize better to real network traffic, and because each has individually achieved good results in the literature~\cite{Silva2026,Belouch2018}. Furthermore, since these datasets include different kinds of benign activity, they also contribute to a more diverse representation of normal traffic. Applying HERA 2.0 to the raw datasets also helped correct existing errors, such as the ones identified in CICIDS-2017~\cite{ZafarIqbalKhan2024,Lanvin2023}.

From this training dataset, six different cyber-attack classes were identified: Bruteforce, \ac{DoS}, Probing, Backdoor, Exploit, Heartbleed, and an additional class for Benign and Background network traffic that is part of a normal computer network. Hyperparameter tuning with 5-fold cross-validation was performed for several common tabular classification models, to identify the most suitable one. Then, the final models were tested with a holdout testing set of network traffic captured during the execution of real cyber-attacks in the deployment environment. Since the models should achieve a robust generalization to normal computer networks, only statistical features related to packet timing, communication statistics, and number of packets in the flows were considered. Due to its adequacy for unbalanced data and consolidation of precision and recall, the macro-averaged F1-score metric was selected to be maximized.

The classification model that achieved the highest macro-averaged F1-score of 85\% was an \ac{XGB} classifier. \ac{XGB} performs gradient boosting using an ensemble of decision trees, with a level-wise growth strategy to split nodes level by level. The key parameters are the number of estimators and the learning rate. The first represents the number of decision trees, whereas the latter controls how quickly the model adapts its weights to the training data. Table~\ref{tab:xgb-params} provides a summary of the optimized hyperparameters that were used in the final deployment.

\begin{table}[h]
\centering
\caption{Summary of XGB configuration.}
\label{tab:xgb-params}
\begin{tabular}{cc}
\toprule
\textbf{Parameter} & \textbf{Value} \\
\midrule
Method & Histogram \\
Loss function & Cross-entropy \\
Number of estimators & 100 \\
Learning rate & 0.3 \\
Feature subsample & 0.9 \\
Maximum depth of a tree & 8 \\
Minimum leaf weight & 1 \\
Minimum loss reduction & 0.01 \\
\bottomrule
\end{tabular}
\end{table}

To perform anomaly detection, the best model was an \ac{IF} detector. \ac{IF} isolates anomalies through an ensemble of decision trees, by repeatedly splitting samples by random values of random features until outliers are segregated from normal observations. This model relies on the contamination ratio of the training set, which must not exceed 50\%. Consequently, the number of samples intended to be anomalies must be lower than the number of remaining samples, otherwise outliers cannot be detected.

This model achieved a slightly lower macro F1 score of 76\%, but it still reliably detected anomalous activity to support the detection of malicious traffic that could not be directly identified as part of a cyber-attack. While \ac{XGB} was tasked with identifying and classifying network flows, \ac{IF} was tasked with identifying patterns and behaviours that deviated from normal flows, in terms of detecting anomalies. Therefore, an anomaly detected solely by this model will not result in an alert, but will be used to highlight when malicious traffic presents anomalous behavior on the network. Table~\ref{tab:if-params} provides a summary of the optimized hyperparameters.

\begin{table}[h]
\centering
\caption{Summary of IF configuration.}
\label{tab:if-params}
\begin{tabular}{cc}
\toprule
\textbf{Parameter} & \textbf{Value} \\
\midrule
Number of estimators & 100 \\
Feature subsample & 0.9 \\
Maximum samples & 1028 \\
Contamination & 0.22 \\
Bootstrap & Disabled \\
\bottomrule
\end{tabular}
\end{table}

After every time window of 60 seconds, the flows are created and exported; the component finds the respective packets' information from the \ac{IDS} export statistics, and creates the connection between packets and flows, based on features such as IP addresses, ports used, and the protocol. As the flow exporter considers bidirectionality, the final state is represented in a single entry. However, the packets in Suricata are considered in several individual messages, spanning from one to multiple requests and responses. As such, the aggregation of packets and flows is done by comparing the start and end of a flow with the start and end of a packet, and its reverse. Furthermore, after a packet is matched with a flow, the identifier value created by Suricata, flow\_id, is used afterwards to match the remaining packets. After the packets are aggregated with the flows, the flow information is sent to both \ac{AI} models, which return their predictions.

Table~\ref{tab:severity} provides an overview of the alert severity levels used for the fusion of alerts for each network flow. Since the signature rules are applied to the packets, the exported flow considers the higher severity detected during its time window. As more components predict that the exported flow is anomalous or part of a cyber-attack class, the final severity level is increased. The final alert is formatted in JSON with a severity level and the detailed prediction of each component, to enable ingestion by the following module and to support more detailed investigations by SOC analysts.

\begin{table}[h]
\centering
\caption{Severity levels by alert fusion from three components.}
\label{tab:severity}
\begin{tabular}{cccccccccccccc}
\toprule
\multicolumn{2}{c}{\multirow{2}{*}{Detection}} & \multicolumn{12}{c}{Severity Level} \\ \cmidrule(l){3-14} 
\multicolumn{2}{c}{} & 2 & 5 & 6 & 6 & 8 & 10 & 10 & 10 & 11 & 12 & 12 & 15 \\ \midrule
\multicolumn{2}{c}{Anomaly} & \cellcolor{mred} \xmark & \cellcolor{mred} \xmark & \cellcolor{mred} \xmark & \cellcolor{mred} \xmark & \cellcolor{mred} \xmark & \cellcolor{mred} \xmark & \cellcolor{mgreen} \cmark & \cellcolor{mgreen} \cmark & \cellcolor{mgreen} \cmark & \cellcolor{mred} \xmark & \cellcolor{mgreen} \cmark & \cellcolor{mgreen} \cmark \\
\multicolumn{2}{c}{Classification} & \cellcolor{mred} \xmark & \cellcolor{mred} \xmark & \cellcolor{mred} \xmark & \cellcolor{mgreen} \cmark & \cellcolor{mgreen} \cmark & \cellcolor{mgreen} \cmark & \cellcolor{mred} \xmark & \cellcolor{mgreen} \cmark & \cellcolor{mgreen} \cmark & \cellcolor{mgreen} \cmark & \cellcolor{mgreen} \cmark & \cellcolor{mgreen} \cmark \\
\multicolumn{1}{c|}{\multirow{3}{*}{\hspace{0.5em} Signature \hspace{0.2em}}} & \multicolumn{1}{c}{\hspace{0.2em} High \hspace{0.5em}} & \cellcolor{mred} \xmark & \cellcolor{mred} \xmark & \cellcolor{mgreen} \cmark & \cellcolor{mred} \xmark & \cellcolor{mred} \xmark & \cellcolor{mred} \xmark & \cellcolor{mgreen} \cmark & \cellcolor{mred} \xmark & \cellcolor{mred} \xmark & \cellcolor{mgreen} \cmark & \cellcolor{mred} \xmark & \cellcolor{mgreen} \cmark \\
\multicolumn{1}{c|}{} & \multicolumn{1}{c}{\hspace{0.2em} Medium \hspace{0.5em}} & \cellcolor{mred} \xmark & \cellcolor{mgreen} \cmark & \cellcolor{mred} \xmark & \cellcolor{mred} \xmark & \cellcolor{mred} \xmark & \cellcolor{mgreen} \cmark & \cellcolor{mred} \xmark & \cellcolor{mred} \xmark & \cellcolor{mred} \xmark & \cellcolor{mred} \xmark & \cellcolor{mgreen} \cmark & \cellcolor{mred} \xmark \\
\multicolumn{1}{c|}{} & \multicolumn{1}{c}{\hspace{0.2em} Low \hspace{0.5em}} & \cellcolor{mgreen} \cmark & \cellcolor{mred} \xmark & \cellcolor{mred} \xmark & \cellcolor{mred} \xmark & \cellcolor{mgreen} \cmark & \cellcolor{mred} \xmark & \cellcolor{mred} \xmark & \cellcolor{mgreen} \cmark & \cellcolor{mred} \xmark & \cellcolor{mred} \xmark & \cellcolor{mred} \xmark & \cellcolor{mred} \xmark \\ \bottomrule
\end{tabular}
\end{table}

\subsection{Analytics Module}

Due to the high volume of alerts that can be raised in an enterprise computer network, it is important to provide a way to summarize them and extract useful patterns and correlations for a security analyst. The analytics module addresses this by automating the aggregation of alert sequences that occur in the same time intervals across different sources, as well as performing an intelligent alert correlation to identify when and where each type of alert is more frequent or rare in a computer network.

The alert sequence extraction is prepared to receive alert logs with varying fields, which are automatically preprocessed to create sequences with relevant statistical features within different time intervals. From the input logs, three main contextual features are extracted: the timestamp of the alert, the specific type of alert according to each detector, and the source host and sub-network of the alert. Based on these fields, the time intervals are analyzed and the unique “type” sequences and unique combinations of “type at source” sequences are created, to enable the correlation between all deduplicated alerts of the same type, and also co-occurring alerts across the same sources.

\begin{figure}[h!]
    \centering
    \includegraphics[width=0.9\linewidth]{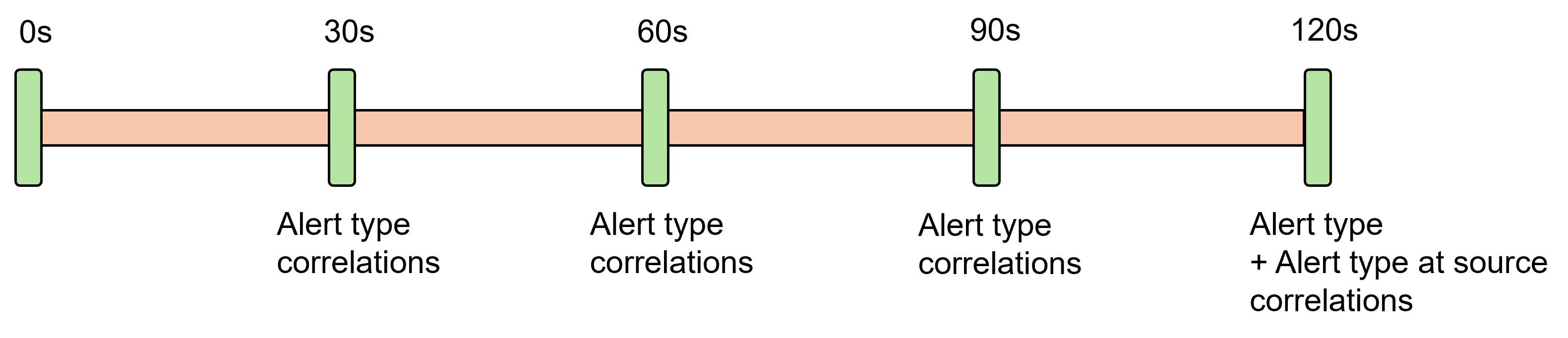}
    \caption{Alert sequence extraction.}
    \label{fig:sequences}
\end{figure}

After a full analysis time window has passed and a batch of multiple sequences is created, alert correlations are extracted using sequential pattern mining algorithms. Figure~\ref{fig:sequences} provides an overview of the creation of one sequence with all deduplicated alert types every 30 seconds, followed by one sequence for each source endpoint over 120 seconds, containing alert types co-occurring at multiple sources during the same time window.

This module executes two parallel sequential pattern mining tasks for "type" and "type at source”, each using an AI algorithm to identify the most frequent patterns and also the rarest patterns that occurred in the full time window. The utilized AI algorithms are two fine-tuned versions of PrefixSpan~\cite{2001_prefixspan}, which is an unsupervised machine learning algorithm that is well-established to identify patterns within sub-sequences of time series data in a computationally efficient manner.

PrefixSpan was fine-tuned according to the contextual features and time intervals used in each task, with minimum support and maximum pattern length hyperparameters. The minimum support is dynamically increased from 1\% to 5\% according to the quantity of input logs within a time window, to improve computational efficiently in the calculation of frequency values. In turn, the minimum support is reduced to one occurrence to find the rarest sequences, stopping the pattern mining execution to use the lowest proportion of occurrences within the time window as their rarity value.

The output of the two parallel sequential pattern mining tasks is postprocessed to obtain the most frequent and most rare patterns and associate each one with the original alerts that occurred during that time window. The mining task for “type” sequences is applied to time windows of 60 seconds, to generate near real-time correlations. In turn, the mining task for “type at source” is applied to windows of 300 seconds to generate more context-aware correlations across multiple sources. These outputs are stored in an interpretable JSON format, and since they are obtained in parallel, there is minimal impact on performance and latency of the previous modules.

Overall, the unified framework provides alert logs from the detection module that are always accessible, as well as output correlations from the analytics module that are generated and stored in near real-time batches, to support intelligent data analytics in a collaborative \ac{SOC}.

\section{Case Study}
\label{sec:case}

\begin{figure}[h!]
    \centering
    \includegraphics[width=\linewidth]{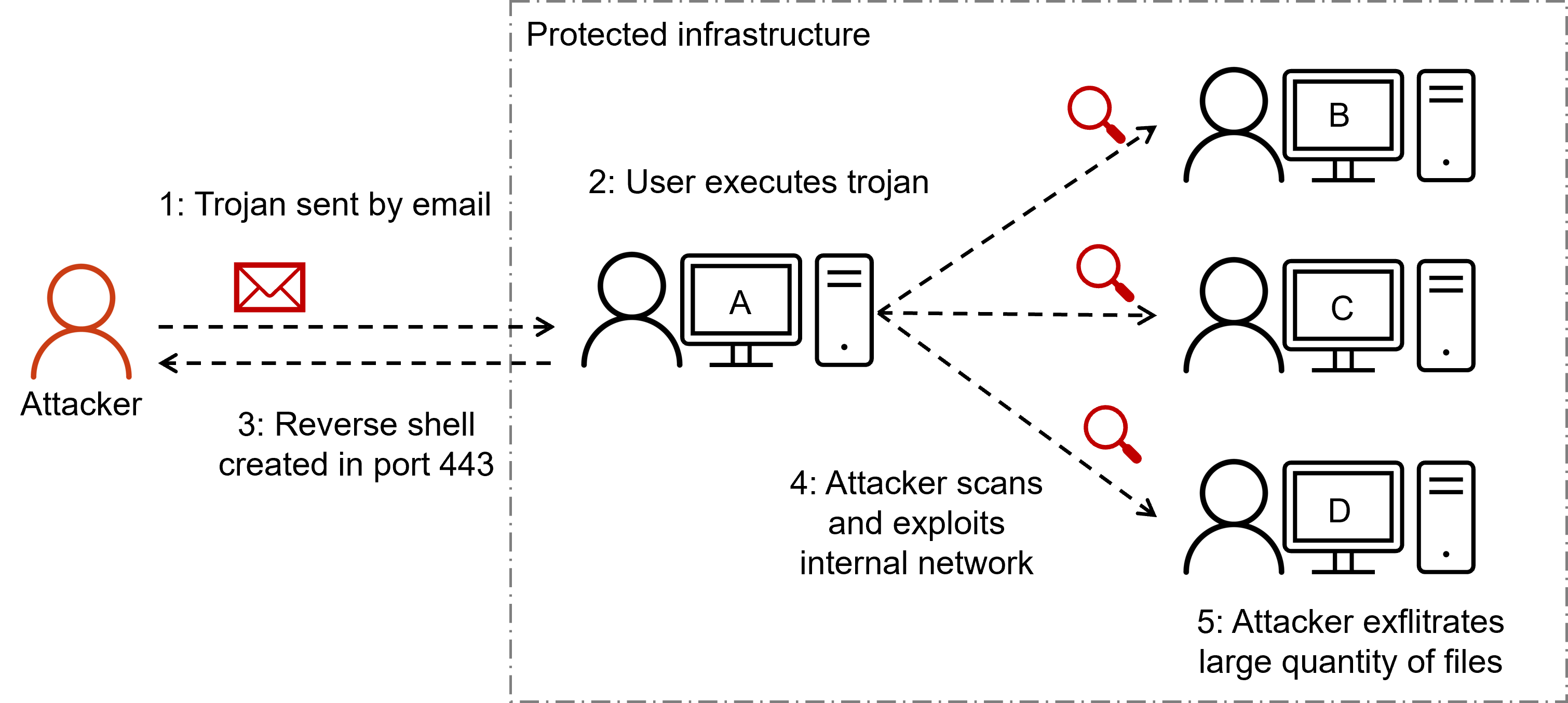}
    \caption{Case study attack scenario.}
    \label{fig:scenario}
\end{figure}

The case study environment includes an attacker machine and a protected infrastructure with multiple worker machines. The infrastructure has a firewall that blocks external connections, to maintain isolation and controlled interaction. Figure~\ref{fig:scenario} provides an overview of an attack vector which begins by sending a spoofed email to one of the workers, containing a Trojan disguised as an application relevant to their work. The worker executes the application, unknowingly establishing a communication tunnel between their machine and the attacker's: a reverse shell using netcat\footnote{\url{https://netcat.sourceforge.net/}} connects over port 443 to bypass the firewall's rules. With this connection established, the attacker performs port scanning to discover active hosts and open ports on the network, then searches the compromised machine for saved credentials, successfully identifying several usernames. Using these credentials to exploit open services on other machines, the attacker locates sensitive data and exfiltrates several files.

\begin{figure}[h!]
    \centering
    \includegraphics[width=\linewidth]{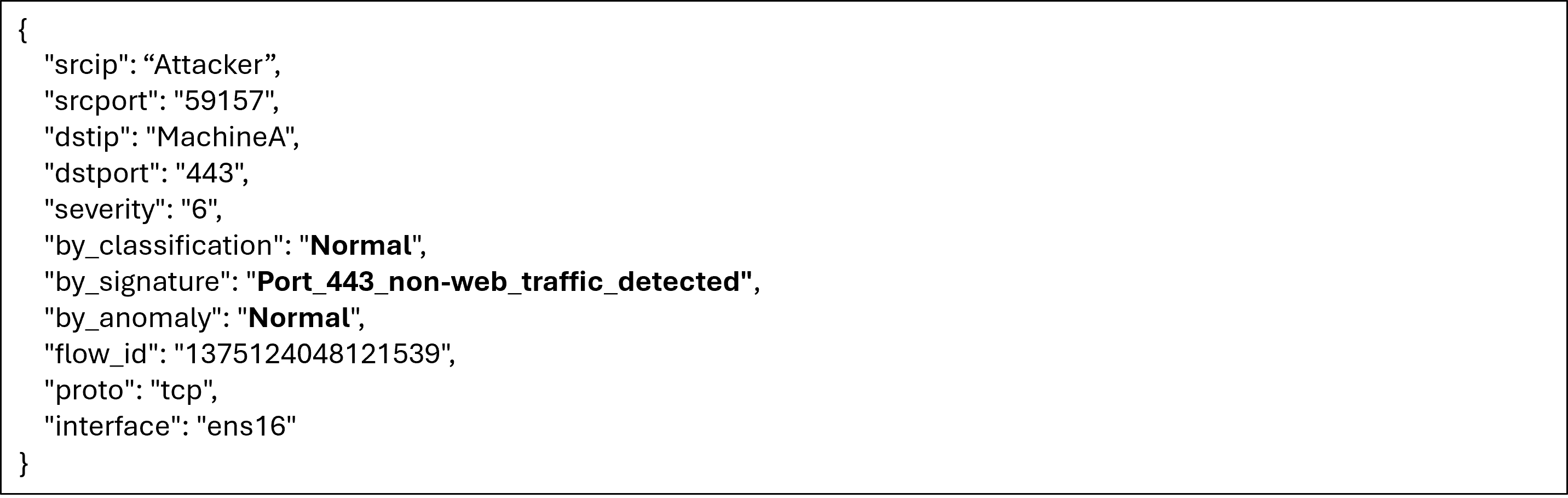}
    \caption{Visualization of an alert created during the creation of a reverse shell.}
    \label{fig:detection_2}
\end{figure}

The router responsible for connecting all machines within the infrastructure to each other and to the internet also mirrors all network traffic to our solution. In addition, a Wazuh\footnote{\url{https://wazuh.com/}} agent was installed on every machine in the infrastructure. Although the email itself is not detected, since the Trojan was designed to bypass both Microsoft Defender and Wazuh's rules, the reverse shell does trigger a rule that flags non-web traffic on port 443, even though it did not trigger any alert from the model. This generates an alert such as the one shown in Figure~\ref{fig:detection_2}, where all the IP addresses were anonymized and simplified, with a severity level of 6, corresponding to a high-priority signature alert.

\begin{figure}[h!]
    \centering
    \includegraphics[width=\linewidth]{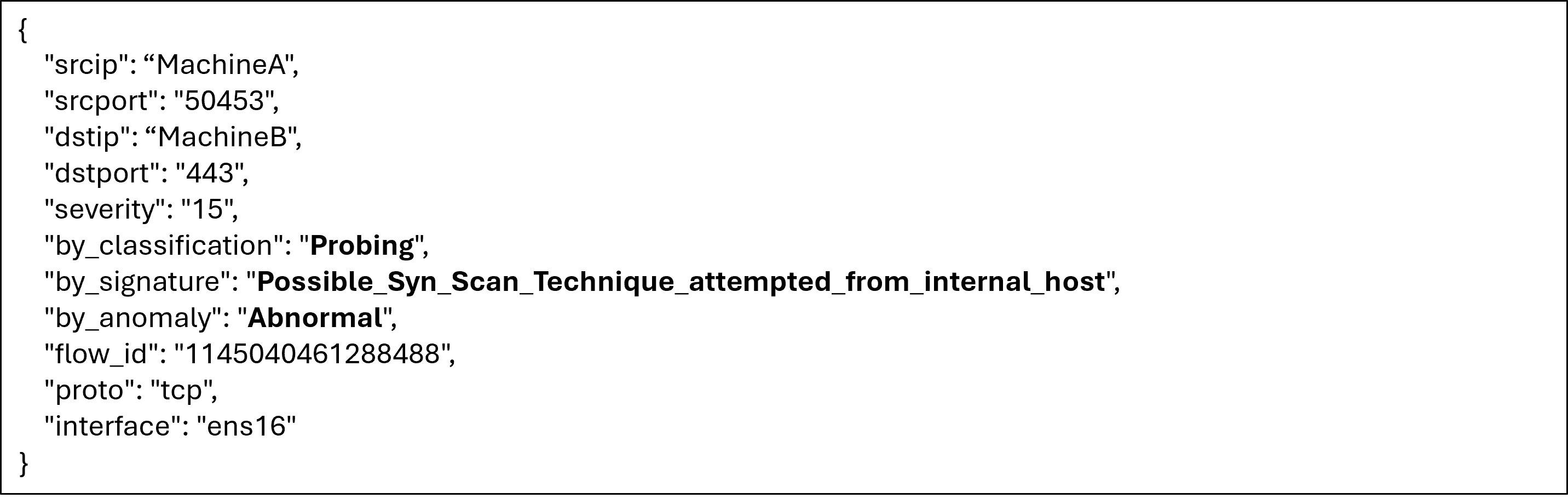}
    \caption{Visualization of an alert created during the scan.}
    \label{fig:detection_4}
\end{figure}

The scan is also detected by the solution; at this stage, every detection method raises an alert, whether classified as a scan from an internal host, probing, or an anomaly, resulting in several alerts of maximum severity, one example of which is shown in Figure~\ref{fig:detection_4}. Finally, the Wazuh agent successfully detected an exploit on the affected machines and raised alerts for the data exfiltration attempt.

\begin{figure}[h!]
    \centering
    \includegraphics[width=\linewidth]{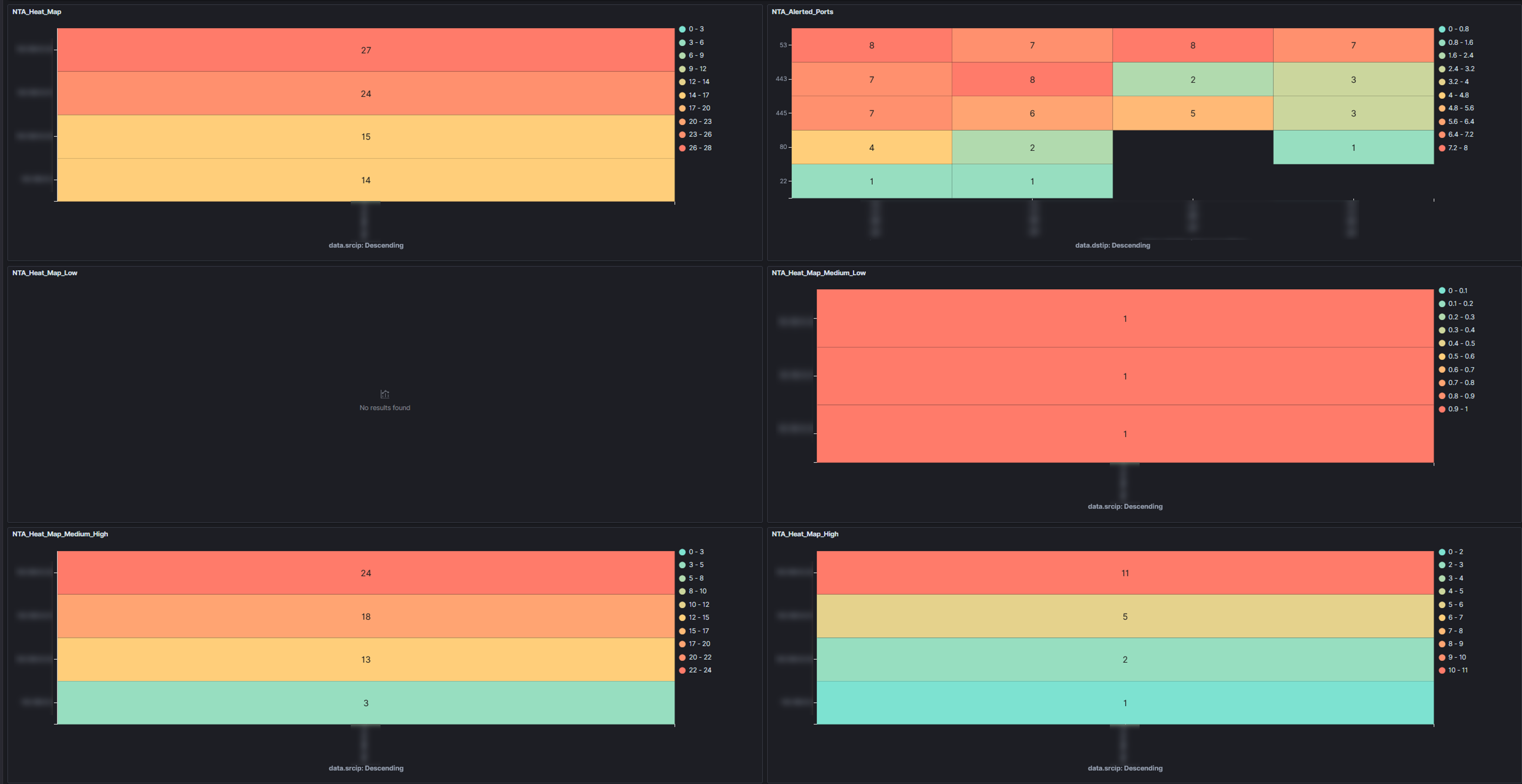}
    \caption{Detection dashboard during case study.}
    \label{fig:detection}
\end{figure}

To better analyze the alerts generated by the proposed solution, a dedicated dashboard was implemented within a Wazuh instance. This presents the alerts in a more structured and intuitive manner, enabling security analysts to monitor them and identify trends over time. This dashboard is shown in Figure~\ref{fig:detection}, filtered to display only the alerts generated during the scan. Several alerts were raised, most of them targeting ports 443 and 53, all originating from MachineA and directed at other machines on the internal network. The majority of these alerts were classified with medium or high severity, indicating that most components were able to successfully identify the attack.

To demonstrate how these alerts are automatically analyzed to extract relevant correlations, consider three sources presenting alert logs: MachineA, MachineB, and MachineC. The repeated logs are deduplicated and summarized in a sequential manner, mixing multiple severity levels in each sequence even if they are not received in order. The logs captured in different machines are organized to accurately represent the flow of events from the network and host, represented by N and H, respectively.

\begin{figure}[h!]
    \centering
    \includegraphics[width=\linewidth]{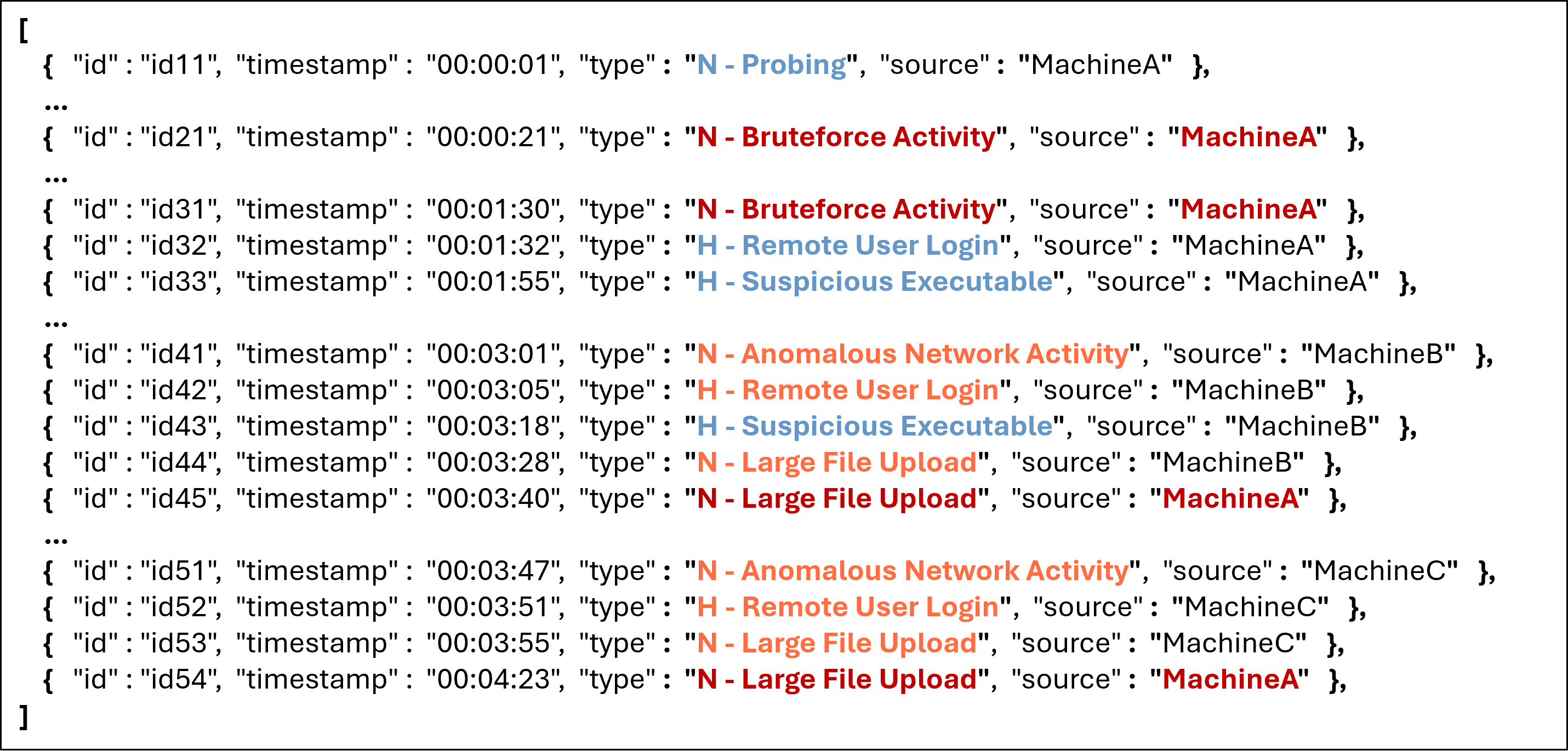}
    \caption{Visualization of alert sequence extraction and preprocessing.}
    \label{fig:preprocessing}
\end{figure}

Figure~\ref{fig:preprocessing} provides an example of the extraction and preprocessing of alert sequences, with “…” representing duplicated alert logs. Each colour highlights different sequences, with red and orange indicating more frequent alerts that occurred multiple times, and blue indicating rarer alerts that only occurred once during each time window. It is important to note that the orange pattern was focused on the “type” of alerts regardless of their source, whereas the red pattern was focused on the “type at source”, be it an alert at a single machine or an alert repeated in multiple machines.

\begin{figure}[h!]
    \centering
    \includegraphics[width=\linewidth]{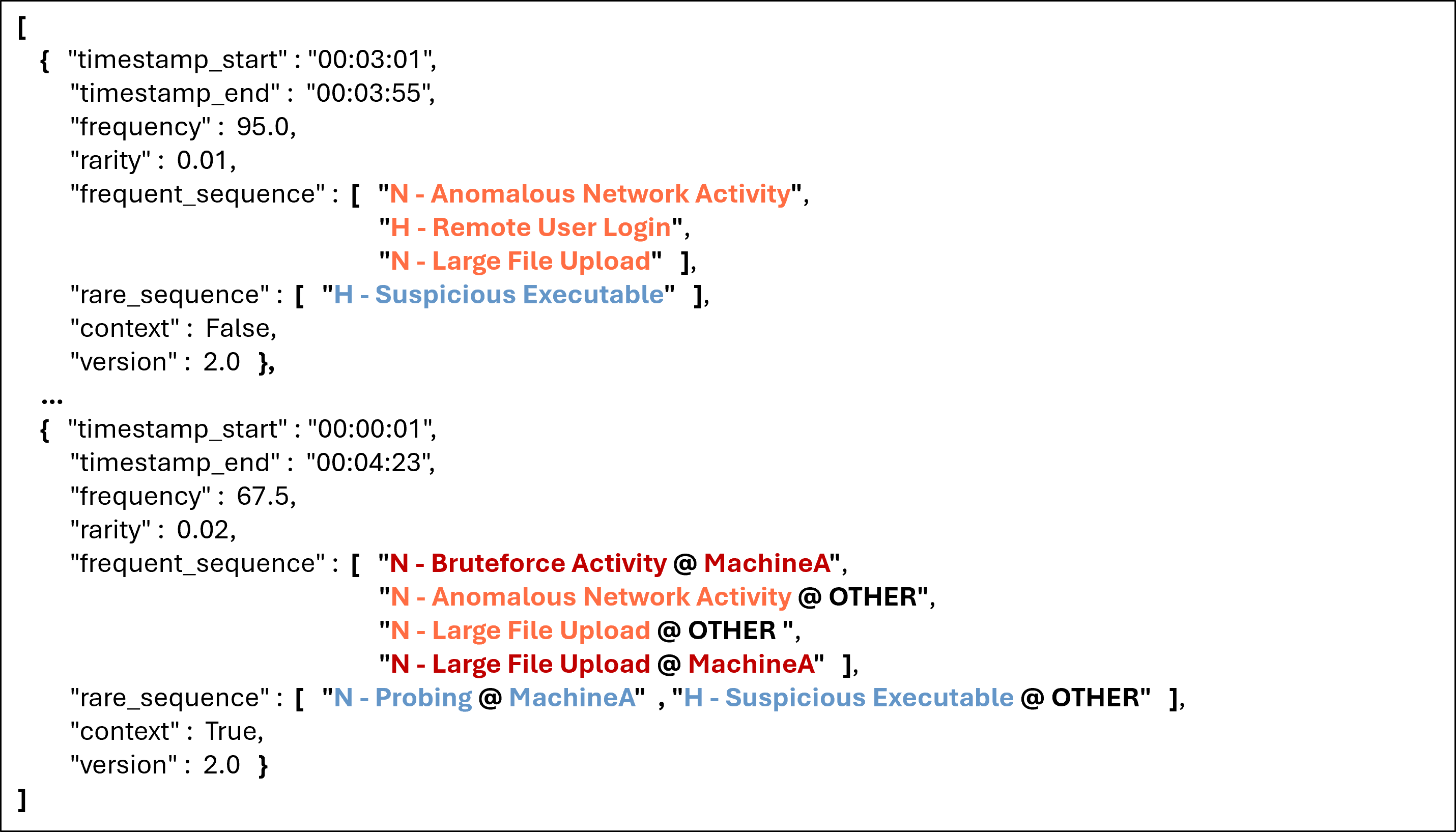}
    \caption{Visualization of alert correlation postprocessing.}
    \label{fig:postprocessing}
\end{figure}

After the analysis, the correlations are postprocessed and stored in an interpretable JSON format, as shown in Figure~\ref{fig:postprocessing}. The first correlation provides the most frequent and rarest “type” sequences within this time window, along with the calculated values for their frequency and rarity. The second correlation provides further context with “type at source” sequences, indicating the host endpoint that is the common element in the sequence. The formatted correlations specify the start and end timestamps of relevant alert logs that were part of this sequence.

It can be observed that the most frequent "type" correlation in this time window highlighted that a large file upload was frequently preceded by a remote user login, which are important steps of the case study attack scenario. Furthermore, the most frequent "type at source" correlation provided important context about the source endpoints where an attack may be occurring. It was automatically identified that there was a common source in alert sequences of this time window, MachineA, because there was a high frequency of file uploads across multiple machines that were followed by similar file uploads in MachineA. These insights, provided in one summarized log instead of hundreds of individual alert logs, enable a security analyst to quickly identify which alerts may be part of the same attack vector.

\begin{figure}[h!]
    \centering
    \includegraphics[width=\linewidth]{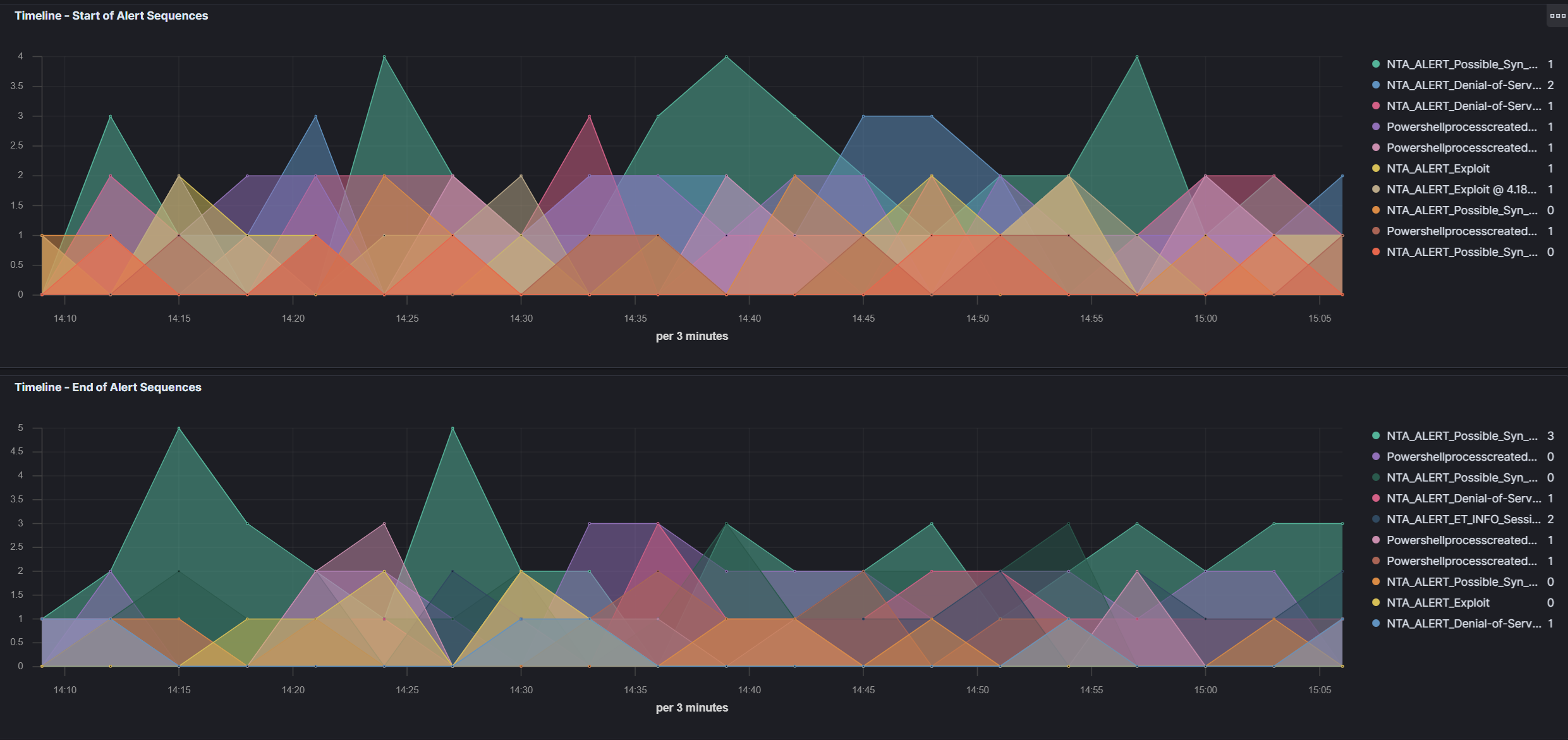}
    \caption{Analytics dashboard during case study.}
    \label{fig:analytics}
\end{figure}

To improve the ease of use of the complete framework, the generated correlations were also represented in a graphical format within the dedicated dashboard. Figure~\ref{fig:analytics} displays a timeline with a summary of the most frequent and rare sequences of alerts that occurred within the selected time window. When a high-severity alert is present at the end of an alert sequence, the correlation module was able to identify common alerts that occurred previously, which may be part of this specific attack vector. In turn, the start of the alert sequences provides an overview of the most common alert logs that may be saturating a SOC.

Overall, the provided alert logs and dashboards enable security analysts to identify common patterns across the high volume of logs recorded in a large and complex network. By combining these patterns with expert knowledge, new filters and signature rules can be developed to further improve detection and adapt them to the deployment environment.

\section{Conclusions}
\label{sec:conclusion}

This work proposed a unified framework to assist \ac{SOC} analysts in the detection of cyber attacks and correlation of alert logs, while maintaining human-in-the-loop verification. The deployment and validation of the proposed solution on an environment with the execution of real cyber attacks have shown to produce good results, as the framework was able to identify attacks with precision, demonstrating its ability to detect and correlate multi-stage attacks effectively while providing security analysts with the structured, actionable insights necessary for timely and informed decision-making.

To continue this work, the scalability of the framework could be further evaluated on machines with more limited computational resources, in order to assess its viability for small and medium-sized enterprises (SMEs), which often lack the infrastructure available to larger organizations. Furthermore, additional components could be integrated within the framework, such as dedicated modules for phishing detection and malware analysis, further broadening its detection capabilities. Finally, it is important to continue exploring and deepening the concepts and mechanisms introduced in this work, particularly with respect to collaborative detection strategies and analyst-in-the-loop workflows.

\begin{credits}
\subsubsection{\ackname} This work was supported by the CYDERCO project, which has received funding from the European Cybersecurity Competence Centre under grant agreement 101128052. This work has also received funding from UID/00760/2025 and support by FCT - Fundação para a Ciência e a Tecnologia, I.P. by project reference 2024.18482.PRT with DOI identifier https://doi.org/10.54499/2024.18482.PRT.

\subsubsection{\discintname}
The authors declare that they have no known competing financial interests or personal relationships that could have appeared to influence the work reported in this paper.

\end{credits}

\bibliographystyle{splncs04}
\bibliography{bibliography}

\end{document}